\documentclass{ptapap}
\usepackage{amsmath, amssymb}
\usepackage{bbold}

\author{Gerardo Urrutia}[CFT]
\author{Agnieszka Janiuk}[CFT]
\affil[CFT]{Center for Theoretical Physics, Polish Academy of Sciences, Al. Lotnikow 32/46, 02-668 Warsaw, Poland}

\title{Following the jet interaction with a post-merger disk outflow}

\begin{document}

\maketitle

\begin{abstract}

Short GRBs are produced by relativistic jets arising from binary NS-NS or NS-BH mergers. Since the detection of the first unambiguous off-axis GRB 170817A, we learned that energy distribution in the jet plays an important role in explaining the GRB emission. The structure and dynamics are modified during the first seconds of the jet interaction with a post-merger environment. Conventional studies often assume this environment as a simple homologous and symmetrically expanding wind. However, post-merger outflows exhibit complex dynamics influenced by the accretion disc evolution. Moreover, the r-process nucleosynthesis influences the thermodynamics and properties of the post-merger neutron-rich environment. In this work, we study the impact of realistic post-merger disc outflow over the jet dynamics at large scales. We find the results are substantially different from the typical model with symmetric homologous wind.

\end{abstract}

\section{Introduction}

%\subsection{What determines the structure of SGRB jets?}
Short gamma-ray bursts (SGRBs) are intense flashes of gamma-rays lasting $t \lesssim 2$~s followed by a multi-frequency, long-lasting, afterglow radiation. The emission of SGRBs results from the propagation of powerful ($E\sim 10^{52}$~erg) relativistic jets into a post-neutron star merger environment, launched by a central engine that consists of a black hole surrounded by an accretion disc \citep{paczynski91}. The SGRB jets acquire their mean physical characteristics during their interaction with the post-merger environment, in particular, the strong outflows coming from the central engine regulate the angular energy distribution of the jet, and its structure \citep[e.g.,][]{Aloy2005,granot2005,granot2018,Murguia2014,Murguia2017,lazzati18,Urrutia2021ShortGRBS,pavan2021,GottliebNakarExpandingMedia2021,Lazzati_2021_mergerEjecta,Nativi2022,Pavan2023}. 

Simulating the complete GRB jet evolution, from the central engine (small scales $\lesssim 10^8$~cm) until its propagation to the interstellar medium (large scales $\gtrsim 10^{13}$~cm) is a very challenging task. Typically, the evolution of GRB jet is solved separately in two regions, at small ($r\lesssim 10^8$~cm) or large scales ($r\gtrsim 10^8$~cm). Small-scale simulations are frequently focused on the physical conditions for jet launching, and the jet evolution is followed during $t\lesssim 1$~s \citep{JaniukJames2022}. However, the implications at large scales are not solved. On the other hand, large-scale simulations assume a simple top-hat jet propagating far away from the central engine. The jet crosses a post merger outflow, but its description is often simplified, notwithstanding the consistency with the previous evolution of jet/outflows at small-scales is washed out.

In this study, we present results of new two-dimensional special relativistic hydrodynamical (SRHD) simulations, of the interaction between a short GRB jet with a realistic post-merger outflow. This study aims to follow this interaction until large scales to determine the impact of the post-merger outflow on the final dynamics of the jet.

\section{Methods}

We use the adaptive mesh refinement \emph{Mezcal} code \citep{decolle12} that employs a second-order shock-capturing scheme. We utilise a 2-dimensional computational box in spherical coordinates $(r,\theta)$ with azimuthal symmetry. The number of cells employed along $r$ direction is $n_r=1\times 10^{4}$, while in $\theta$ direction is $n_\theta = 100$. The inner boundary is fixed at $r_{\rm min}=3\times 10^{8}~$cm and the outer radius is located at $r_{\rm max} = 1.2\times 10^{11}~$cm. The maximum resolution of the grid is $\Delta r\approx 1.49 \times 10^{6}\,$cm and $r_{\rm min} \Delta \theta \approx 1.17 \times 10^{6}\,$cm, in $r$ and $\theta$, respectively. Our simulations were performed over an integration time of $t=4~$s.

We fill the computational box with a constant and static ambient density environment $\rho_a=10^{-5}~$g cm$^{-3}$. It does not affect the evolution of the jet and post-merger the outflow. At the injection radius $r_{\rm inj}$ we impose the initial conditions trough the primitive variables $(\rho,v,p)$. Below we explain the imposition of the post-merger outflow and the jet, respectively.

\begin{figure}
    \centering
    \includegraphics[width=\textwidth]{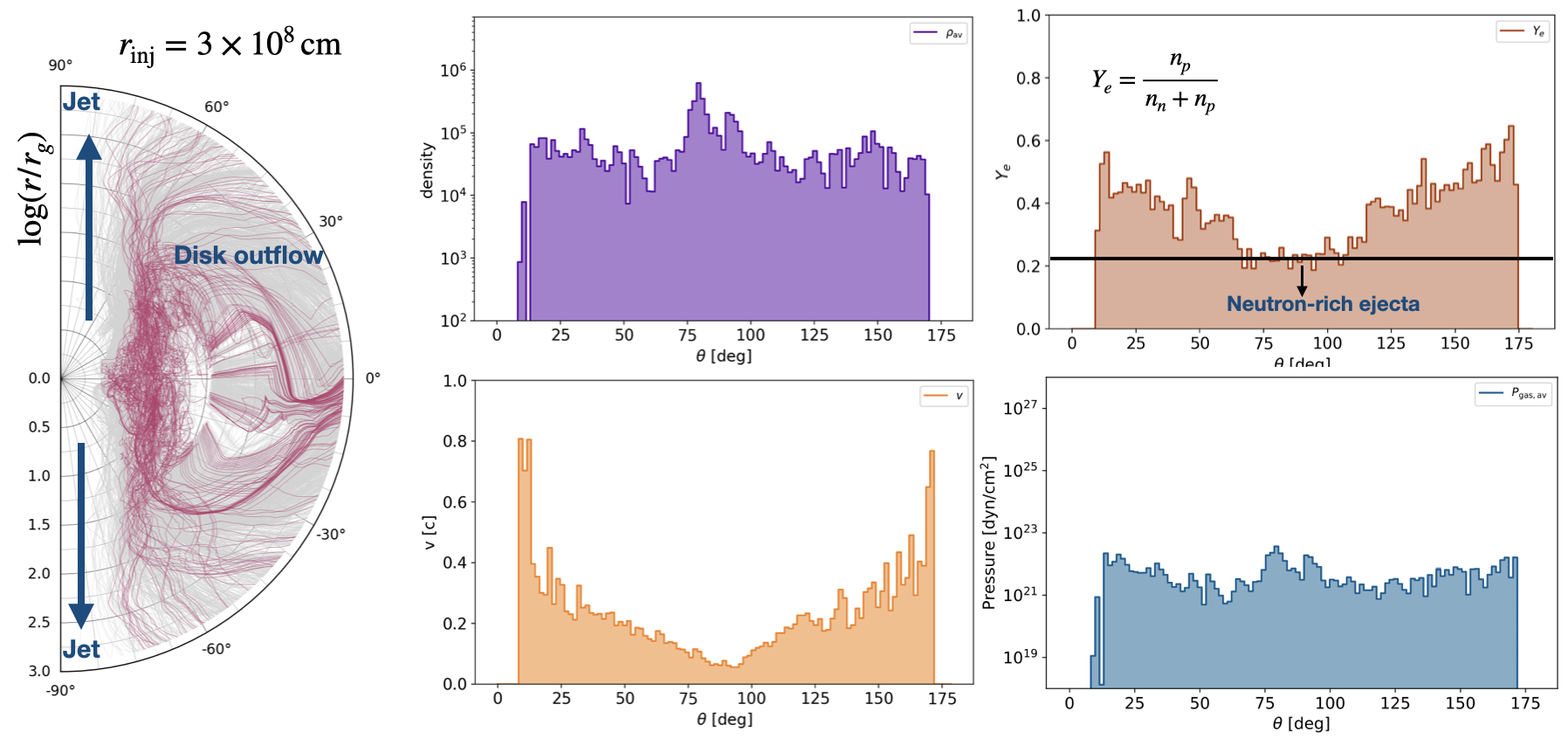}
    \caption{The distribution of the post-merger disk wind outflow at $r_{\rm inj}=3\times 10^8~$cm. It represents the initial conditions for the large-scale simulation.}
    \label{fig:init_conds}
\end{figure}

The post-merger outflow is based on the model M2.65-0.1-a0.9 from \cite{Nouri_2023}, resulting from the evolution of an accretion disk formed after the NSNS merger. This model was performed in the General Relativistic Magneto Hydrodynamic (GRMHD) HARM-COOL code \citep{Gammie_2003,Janiuk:2013,Janiuk2019}, considering a realistic nuclear equation of state and neutrino cooling. The evolution of thousands of fluid parcels is followed by tracers which are post-processed using the SkyNet nuclear reaction network \citep{Lippuner_2017}. We extract the pressure influenced by nuclear burning by employing an inversion of the Helmholtz equation \cite{TimmesArnett1999ApJS}. At $r_{\rm inj}$, we remap the distributions of density, pressure and velocity, which are stored in tracer particles (Figure \ref{fig:init_conds}). The injection of wind covers the timescale $t_{\rm inj}\lesssim 0.4~$s based on the integration time of the previous GRMHD simulation. For greater times, we assume that time-averaged distributions can be approximated as $\rho(t>t_{\rm inj})\propto \rho (t/t_{\rm inj})^{-5/3}$. This post-merger outflow is denoted as ``NSNS'' in Table \ref{tab:list_models}.

To mitigate the impact of disk wind effects and make a comparison with the NSNS case, an alternative post-merger outflow is assumed as a homologously expanding wind \citep[e.g.,][]{Lazzati_2021_mergerEjecta}. It is described by 
$\rho (r) = \rho_* e^{-r/r_{\rm inj}}$, where we fixed $\rho_* = 10^{-2}\,M_{\odot}$~s$^{-1} /\, (4\pi r_{\rm inj}^2\,10^{-1}c)$. In Table \ref{tab:list_models}, this case is denoted as ``SPH''. 

For the jet implementation, we use the last snapshot of the GRMHD simulation to estimate the luminosity of the jet as
\begin{equation}
    L_j = 2\pi \int_{0}^{\theta_j} \left(  \Gamma\rho h c^2 - p - \Gamma\rho c^ 2 \right) v_r\, r_{\min}^2\sin\theta\, d\theta.
    \label{eqn:jet_lum}
\end{equation}
Here, the jet opening angle is $\theta_j=15^\circ$ and the contribution of magnetic energy\footnote{An example of the estimation of jet luminosity, taking into account the contribution of magnetic energy, is doing by \cite{James_2022, JaniukJames2022}.} is omitted. The density and pressure of the jet are based on a strong shock condition \citep[e.g.,][]{urrutia22_3D}, given respectively by
\begin{equation}
    p_j = \frac{\rho_j c^2}{4} \left( \frac{\Gamma_\infty}{\Gamma_j} -1 \right),  \qquad {\rm and} \qquad   \rho_j = \frac{L_j}{\Gamma_\infty \Gamma_j v_j c^2 \Delta S},
    \label{eqn:high_jet_pressure}
\end{equation}
where $\Delta S = 4 \pi \left( 1-\cos \theta_j \right) r_j^2 $, and the terminal Lorentz factor is assumed $\Gamma_\infty=100$. The injection time for the jet $t_j$ is restricted by the averaged mean lifetime of the central engine \citep{1998ApJ...494L..53K,2002ApJ...577..893L,2008ApJ...675..519J}, approximated by $t_j = m_d/\langle \dot{M}_{\rm in} \rangle$.

We summarise the main parameters used for this study in Table \ref{tab:list_models}.

\begin{table}
\centering
\begin{tabular}{|c||c|c|c|c||c|c|c|c|} 
 \hline
 Model & $L_j$ & $\theta_j$& $\Gamma_j$ & $t_j$ & $M_{\rm BH}$ & $a$ & $M_{\rm disk}$  & $\dot{M}_{\rm out}$ \\  
 &[erg~s$^{-1}$]&[$^\circ$]&&[s]&&&&[$M_\odot$~s$^{-1}$] \\
 \hline \hline

 NSNS & $1.4\times 10^{50}$ & $15^\circ$& $7.4$ &$1.57$ & $2.65$& $0.9$&$0.10276$ & $3.27\times 10^{-2}$  \\  
 SPH & $1.4\times 10^{50}$ & $15^\circ$& $7.4$ &$1.57$ & -- & -- & -- &-- \\

\hline

 \end{tabular}
    \caption{The main parameters of the jet and the disk wind.}
    \label{tab:list_models}
\end{table}

\section{Results}

\begin{figure}
    \centering
    \includegraphics[width=\textwidth]{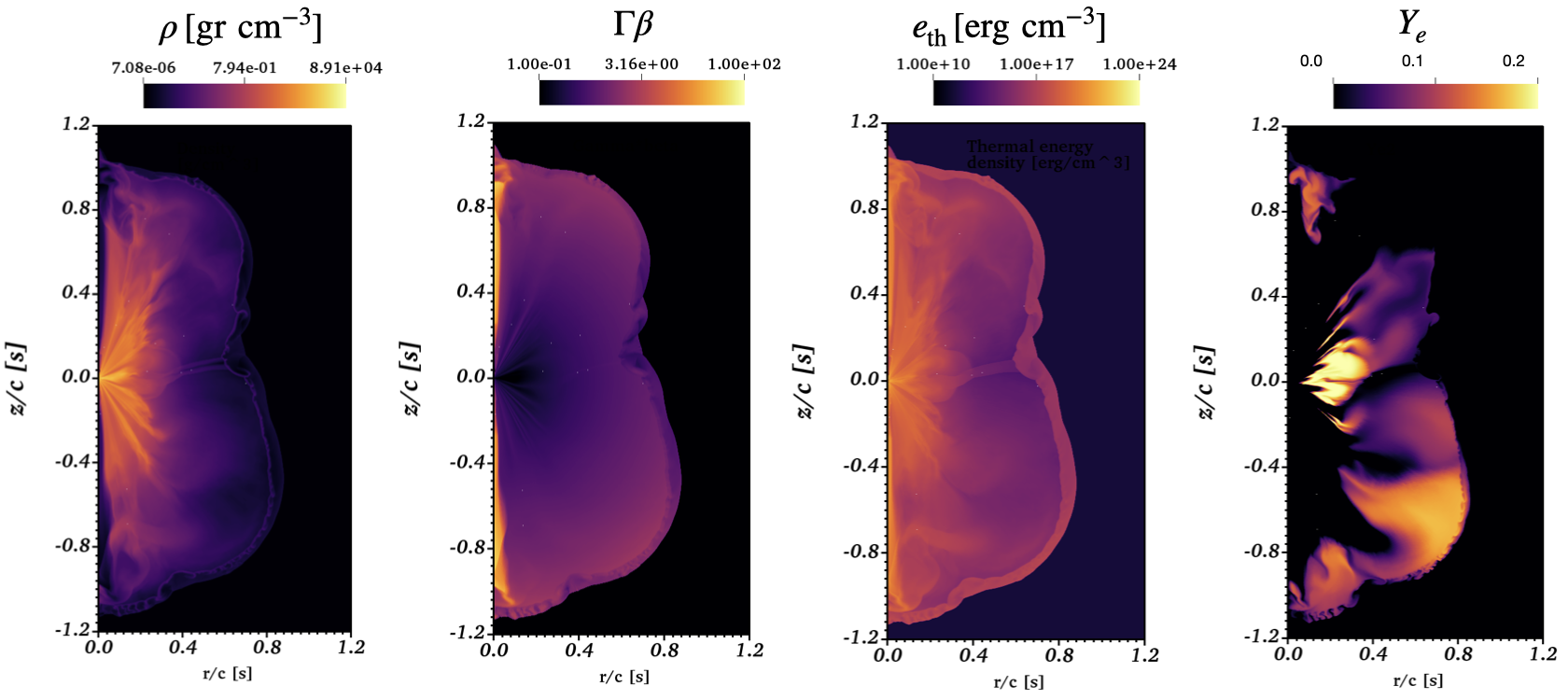}
    \caption{The snapshot at $t=1.2~$s. The maps from left to right show, respectively: density $\rho$, velocity $u=\Gamma\beta$, thermal energy $e_{\rm th}$, and the passive scalar that represents the distribution of electron fraction $Y_e$. }
    \label{fig:maps}
\end{figure}

In Figure \ref{fig:maps} we present maps taken at $t=1.2~$s and showing the distributions of density $\rho$, velocity $u=\Gamma\beta$, and thermal energy $e_{\rm th}$. In addition, the passive scalar represents the electron fraction distribution, $Y_e$. The density distribution shows that the most heavy material remains close to the central region, and a stratification pattern is not obvious. The velocity map confirms that the heavy material is slower, and remains at the inner regions. On the other hand, the jet region is contrasted by the velocity, and recollimation shocks are formed. A fast shell is created in front of the wind, and its velocity decreases in regions close to the equator. 

\begin{figure}
    \centering
    \includegraphics[width=\textwidth]{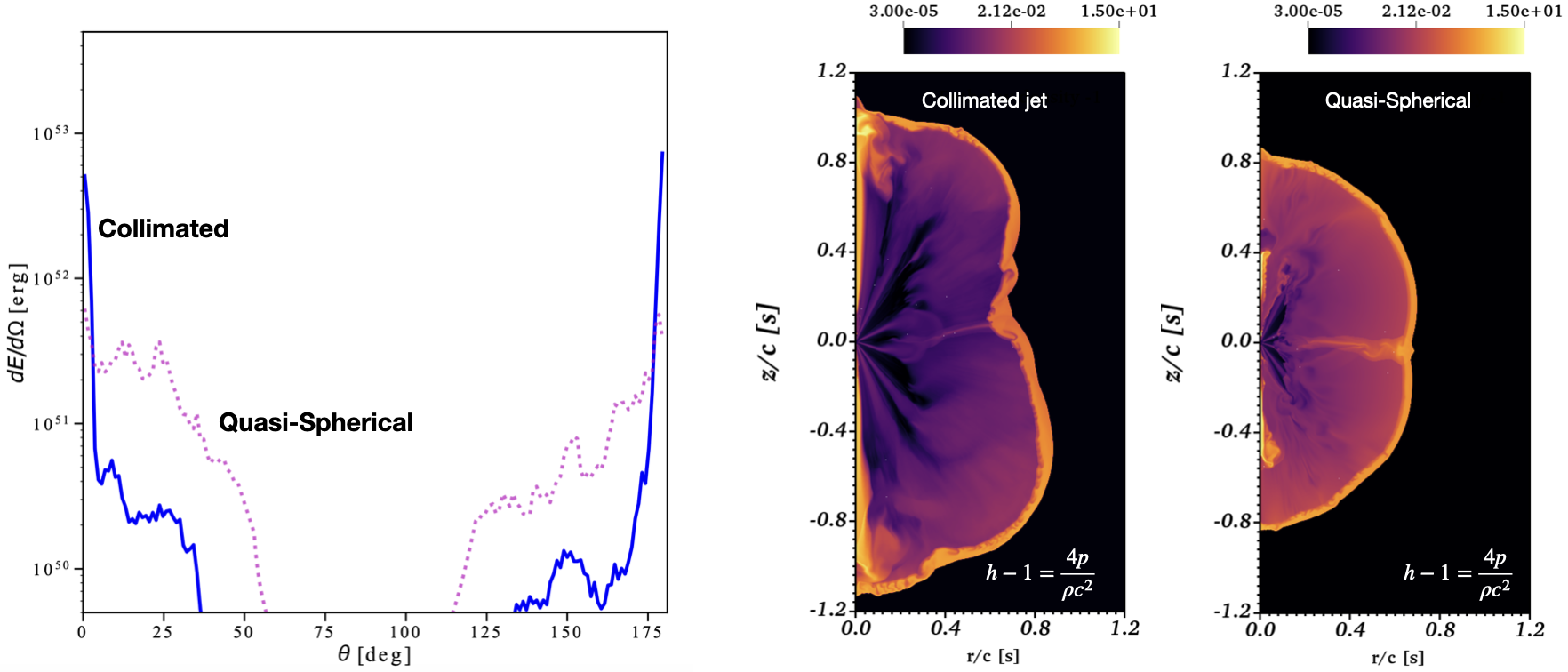}
    \caption{The snapshot at $t=1.2~$s. The left panel shows the angular distribution of thermal energy in the jet and wind. The enthalpy maps are plotted with color scale, for the post-merger disk wind (middle panel) and homologous spherical wind (right).}
    \label{fig:maps_2}
\end{figure}

The thermal energy map shows the most representative regions of interaction between the jet and the wind. A more detailed distribution\footnote{Both contributions, thermal and kinetic will be discussed in Urrutia et al. (in prep).} is showed in Figure \ref{fig:maps_2}. There is a clear distinction depending on the post-merger environment. For example, jets remain collimated by the realistic post-merger outflow, while more spreaded jets expand into spherical homologous wind. The color maps of enthalpy in Figure \ref{fig:maps_2} show the transition regions, which are heated due to shocks or abrupt changes in the velocity. In the post merger environment, the transition regions are highly contrasted, due to strong shock heating. On the other hand, in the homologous case almost all the sphere remains uniformly thermalized. 

In general, the distributions are not symmetrical. 
%It can be We present asymmetric distributions of the system, in general, is non symmetric. 
It should be noted, that the electron fraction distribution suggests an asymmetric outflow of less neutronized material towards intermediate latitudes, where some lighter elements ($2^{nd}$ peak of r-process) can be synthesized. The more neutronized material is expanding rather towards the equatorial plane, and distribution of 
the $3^{rd}$ peak elements is more symmetric.
%which flag each region of the material and it is possible to follow its mixing with different regions. 

\section{Discusion and Conclusions}

We followed the interaction between short GRB jets and a realistic post-merger outflow. We introduce jets with similar characteristics ($\theta_j$, $L_j$, $\Gamma_j$) into distinct post-merger environments: one obtained from the GR MHD simulations of accretion disk wind which is the site of r-process nucleosynthesis (realistic case), and the other model where the environment is described as a stratified homologously expanding spherical wind (simplified case).

At large scales ($r \gtrsim 10^8~$cm), we observe different patterns of energy distribution between these two cases. The jet propagating in a homologous environment exhibits a similar behaviour reported in previous works \citep[][e.g.,]{Aloy2005,Duffell2015,duffell18,Hamidani2020,Urrutia2021ShortGRBS,GottliebNakarExpandingMedia2021,Murguia_Berthier_2021}, where the energy is spreaded smoothly. However, the impact of the disk wind exhibits distinct effects, maintaining the collimation of the jet.

A novel contribution of this work involves careful consideration of the initial pressure injection, influenced by nuclear burning resulting from the on-going $r-$process. This pressure is crucial for the balance between the cocoon expansion and jet collimation.

Despite meticulous wind injection techniques, the jet was assumed as a continuous top-hat (whitout structure), where $L_j$ and $\Gamma_j$ were imposed without angular distribution. However, the jet structure can be partially deleted by the strong interaction with the post-merger environment \citep{Nativi2022,urrutia22_3D}. 

Clear delineation of distributions at larger scales remains essential. In conclusion, our findings underscore that consistent differences of thermodynamical properties of post-merger outflows significantly and abruptly influence the jet propagation, establishing it as a critical component in short GRB modelling.

\acknowledgements{This work was supported by the grant 2019/35/B/ST9/04000 from Polish National Science Center. This research was also supported by PLGrid Infrastructure.}

\bibliographystyle{ptapap}
\bibliography{urrutia}

\end{document}